\documentclass[conference]{IEEEtran}
\IEEEoverridecommandlockouts
\usepackage{amsmath,amsthm,amssymb,amsfonts}
\usepackage{bbm}
\usepackage{algorithm}
\usepackage{algorithmic}
\usepackage{graphicx}
\usepackage{textcomp}
\usepackage{xcolor}
\usepackage{flushend}

\usepackage[numbers]{natbib}

\usepackage{subfig}
\usepackage{svg}

\usepackage[normalem]{ulem}

\def\BibTeX{{\rm B\kern-.05em{\sc i\kern-.025em b}\kern-.08em T\kern-.1667em\lower.7ex\hbox{E}\kern-.125emX}}

\begin{document}

\title{VeriRAG: A Retrieval-Augmented Framework for Automated RTL Testability Repair}

\author{
    \IEEEauthorblockN{Haomin Qi$^{1,2,\dagger,\ddagger}$, Yuyang Du$^{1,\dagger}$, Lihao Zhang$^{1}$, Soung Chang Liew$^{1,*}$, Kexin Chen$^{1}$, Yining Du$^{1,3,\ddagger}$}
    \IEEEauthorblockA{$^1$The Chinese University of Hong Kong, $^2$University of California San Diego, $^3$Hong Kong University}
    \IEEEauthorblockA{h.chee@link.cuhk.edu.hk, \{yuydu, zl018, soung\}@ie.cuhk.edu.hk, kxchen@cse.cuhk.edu.hk,  yiningdu@connect.hku.hk}
    \vspace{-4.5em}
    \IEEEauthorblockA{\thanks{$^\dagger$H. Qi and Y. Du contributed equally to this work.}}
    \IEEEauthorblockA{\thanks{$^\ddagger$Work conducted during internship at IE Department, CUHK.}}
    \IEEEauthorblockA{\thanks{$^*$S. C. Liew (soung@ie.cuhk.edu.hk) is the corresponding author.}}
    \IEEEauthorblockA{\thanks{The work was partially supported by the Shen Zhen-Hong Kong-Macao technical program (Type C) under Grant No. SGDX20230821094359004.}}
}

\maketitle
\begin{abstract}
Large language models (LLMs) have demonstrated immense potential in computer-aided design (CAD), particularly for automated debugging and verification within electronic design automation (EDA) tools. However, Design for Testability (DFT) remains a relatively underexplored area. This paper presents \textit{VeriRAG}, the first LLM-assisted DFT-EDA framework. VeriRAG leverages a Retrieval-Augmented Generation (RAG) approach to enable LLM to revise code to ensure DFT compliance.  VeriRAG integrates (1) an autoencoder-based similarity measurement model for precise retrieval of reference RTL designs for the LLM, and (2) an iterative code revision pipeline that allows the LLM to ensure DFT compliance while maintaining synthesizability.  To support VeriRAG, we introduce \textit{VeriDFT}, a Verilog-based DFT dataset curated for DFT-aware RTL repairs. VeriRAG retrieves structurally similar RTL designs from VeriDFT, each paired with a rigorously validated correction, as references for code repair. With VeriRAG and VeriDFT, we achieve fully automated DFT correction -- resulting in a 7.72-fold improvement in successful repair rate compared to the zero-shot baseline (Fig. \ref{fig:success_rates_2} in Section \ref{sec:Experiments}). Ablation studies further confirm the contribution of each component of the VeriRAG framework. We open-source our data, models, and scripts at \textcolor{blue}{https://github.com/HarminChee/VeriRAG}.

\end{abstract}


\section{Introduction}
\label{sec:introduction}
Recent advancements in large language models (LLMs) have demonstrated significant potential in enhancing computer-aided design (CAD) workflows, particularly in automated debugging and verification of hardware description languages (HDLs) \cite{blocklove2024evaluating,firouzi2025chipmnd}. However, despite the growing interest in LLM-empowered CAD tools, the domain of design for testability (DFT) remains comparatively underexplored. Unlike functional verification, which establishes behavioral correctness, DFT at RTL aims to secure observability/controllability and test access properties (e.g., scan readiness, safe clock/reset handling) that enable robust ATPG and manufacturing test.

As systems scale to millions of transistors and face tight timing, yield, and power constraints, addressing testability at the register-transfer level (RTL) stage becomes imperative -- resolving DFT issues early enables synthesis tools to optimize area and facilitate reuse without requiring repeated fixes \cite{wang2006vlsi, thakur2022current}. Moreover, many system-on-chip (SoC) designs integrate IP cores from diverse sources, making RTL-level testability crucial for meeting fault coverage goals and minimizing iterative cycles of synthesis, verification, and automatic test pattern generation (ATPG). In practice, engineers still rely heavily on time-consuming manual processes to ensure DFT compliance. There is an urgent need for automated approaches to perform DFT-oriented HDL code repairs\cite{alsaqer2024potential, abdollahi2024hardware}.

We propose \textit{VeriRAG} --- a retrieval-augmented generation (RAG) framework for efficient LLM-based RTL repair for DFT compliance --- along with an associated DFT dataset named \textit{VeriDFT}. VeriRAG incorporates (1) an autoencoder trained by the contrastive learning method to identify DFT-related hardware similarities, and (2) an iterative error correction pipeline guided by compiler diagnostics and DFT error reports. VeriDFT is curated from an open-sourced Verilog dataset \cite{GitDataset}, augmented by our manual annotations to capture classic DFT errors in Verilog HDLs.

Given a Verilog code snippet potentially requiring DFT-compliance code revisions, VeriRAG analyzes the hardware structure of the RTL design and retrieves the most similar example from the VeriDFT dataset based on both hardware structure and DFT errors. The retrieved example, along with our manually edited solution for DFT compliance, serves as a tailored in-context reference to guide the LLM’s correction of the given Verilog code snippet. Each reference in the library is paired with a validated fix, which mitigates the risk of misleading edits from mismatched or unverified exemplars.

Importantly, VeriRAG’s error correction mechanism iteratively refines the RTL design by leveraging compiler diagnostics and DFT error reports generated by electronic design automation (EDA) tools. This iterative process ensures that the modifications effectively address DFT concerns while preserving the design’s synthesizability and logical integrity. A repair is accepted only if (i) the design remains synthesizable, (ii) no DFT violations are reported, and (iii) a final logic equivalence check (LEC) passes against the original.

To validate its effectiveness and robustness across diverse language models, we evaluate VeriRAG using several state-of-the-art LLMs, including GPT-o1, Grok-3, GPT-4o, Claude-3.7-Sonnet, and Gemini-2.5-Pro. Among these models, GPT-o1 achieves the highest rate of success for DFT error correction --- a successful correction is defined as eliminating DFT errors while retaining synthesizability and logical equivalence. Given the inherent complexity of this task, the overall success rate remains modest, with GPT-o1 achieving a 53.76\% success rate on the test set. Despite this, the result represents a significant milestone, marking a 7.72-fold improvement over the baseline success rate of 6.96\%, achieved by the same LLM using naive prompting strategies (see Fig. \ref{fig:success_rates_2} in Section \ref{sec:Experiments}). Comprehensive ablation studies were conducted to validate the necessity of each component within VeriRAG.

As an initial effort in the filed, this paper makes the following contributions:
\begin{itemize}
    \item \textbf{\textit{VeriRAG} Framework for DFT-Compliance Code Repair}: We put forth VeriRAG, a RAG-based pipeline that leverages an autoencoder-driven similarity model and an iterative, compiler-guided revision loop to improve LLM performance on DFT-oriented Verilog repair tasks.
    \item \textbf{Domain-Specific Dataset \textit{VeriDFT}}: To support VeriRAG, we construct VeriDFT, the first RTL design dataset that not only catalogs representative DFT error patterns but also supplies corresponding manually validated corrections for each example. VeriDFT fills a critical gap in hardware testability research, offering high-quality data tailored for DFT-driven retrieval and repair. 
    \item \textbf{Benchmarks and Open-Source Contributions}: We conduct extensive evaluations of VeriRAG, establishing the first benchmarking baseline for LLM-based DFT error corrections at the RTL stage. To foster reproducibility and future research, we release all related resources, including (1) the reference implementation of VeriRAG, (2) the VeriDFT dataset, and (3) EDA scripts for compiler diagnostics, DFT error reporting, and logic equivalence check (LEC).
\end{itemize}

\section{Background and Related Works}
\label{sec:RelatedWork}
\subsection{LLM for CAD} 
LLMs are increasingly being used in CAD, supporting tasks ranging from high-level synthesis to RTL bug detection. Recent studies have shown that LLMs can assist with targeted design improvements, automate documentation, and even interface with EDA tools for power and timing analysis based on user inputs \cite{chipchat, VerilogEval, DeepRTL, CodeV, du2024VTC, Verigen}. However, most of these efforts have focused on functional correctness, such as verifying data paths or refining structural descriptions, while largely neglecting the critical aspect of testability \cite{zhao2025vrank}. Our work addresses this gap by targeting \emph{DFT compliance} at the RTL level, where success requires not only compilable edits but also the elimination of DFT violations and preservation of logical equivalence. In contrast to repository- or function-oriented frameworks (e.g., AutoVCoder; RTLRepoCoder) that primarily pursue functional completion or synthesis-aligned correctness, our setting is testability-driven and closes the loop with tool-based DFT diagnostics and a final logic equivalence check (LEC).

\subsection{DFT at RTL level}
Conventional DFT techniques --- such as scan-chain insertion \cite{Ref8}, boundary scan \cite{Ref9}, and built-in self-test \cite{Ref10} --- primarily operate at the gate level or during physical implementation. While these methods automate many aspects of test generation and reduce manual effort in late design stages, they do not address defects introduced at the RTL stage. Consequently, designers must rely on linting and static analysis tools to identify code-quality issues. The systematic identification and correction of DFT-related errors, particularly those involving internal clocking or asynchronous resets, remains labor-intensive and requires significant expertise \cite{Ref11}.From a process perspective, gate-level DFT insertion cannot reliably repair RTL-origin violations (e.g., internally generated clocks or uncontrollable resets); therefore, catching and fixing such patterns \emph{early at RTL} is crucial for avoiding repeated synthesis–ATPG iterations and schedule risks. We focus on four representative RTL-level DFT patterns: ACNCPI (asynchronous set/reset not controllable from primary inputs), CLKNPI (internally generated or gated clocks), CDFDAT (a clock net used as data), and FFCKNP (flip-flop–driven clocks).

\subsection{RAG for RTL design}
RAG combines generative modeling with the retrieval of relevant external knowledge to enhance performance. When employing RAG for our purpose, code templates and bug-fix repositories can be leveraged to guide language model outputs, improving both the precision and correctness of LLM-assisted code repairs \cite{Ref12, qi2025graphcuesdnconfigurationcode}. For RTL design, previous retrieval-based methods \cite{Ref13} have shown that proactively identifying and reusing well-validated IP and design blocks can significantly improve development efficiency. However, applying RAG to DFT repair introduces a unique challenge: defining and measuring ``similarity” for the precise retrieval of the most relevant reference from the knowledge base. In particular, DFT cues (clocks, resets, scan enables) are \emph{sparse and topology-dependent}, so text-level similarity is insufficient. We therefore adopt a structure-aware retrieval formulation that quantifies hardware similarity with respect to DFT-relevant connectivity, enabling precise selection of validated reference–fix pairs within the RAG loop.

\subsection{Deep Learning for Hardware and DFT Error Analysis}
Deep learning has recently become a powerful tool for hardware analysis, achieving notable results in power modeling \cite{Ref14}, defect localization \cite{Ref15}, and timing optimization \cite{Ref16, Ref17}. Most existing methods cast hardware analysis as classification or regression tasks, such as anomaly detection, layout optimization, or performance prediction. In the context of DFT, a complementary line of work learns embeddings of RTL structures so that designs sharing DFT-relevant topologies are close in a latent space; similarity can then be computed (e.g., via cosine distance) to retrieve a small number of validated references that guide subsequent edits. Our study situates DFT-oriented similarity learning within this landscape and employs it strictly as a \emph{retrieval} component, while the actual code repair is governed by compiler/DFT diagnostics and LEC-based acceptance.

\section{The VeriDFT Dataset}
\label{sec:Dataset}
\subsection{Data Cleaning and Data partition} \label{sec:DatasetA}
Our VeriDFT dataset builds upon a publicly available data collection containing 108,971 Verilog code samples \cite{GitDataset}. The original dataset contains substantial DFT-irrelevant content that may impair the effectiveness of downstream analysis. Comprehensive data cleaning is required. The preprocessing steps outlined below address issues such as non-functional files, extraneous modules, and compilation errors.

First, we filter out Verilog files without meaningful logic functionality, such as testbenches and module wrappers. While these files are essential for simulation, they do not contain circuit designs relevant to DFT analysis. Second, we notice that many designs depend on pre-defined low-level modules or built-in IP cores provided by EDA tools (e.g., Xilinx IP cores in Vivado). Although our EDA tools report errors due to missing IP cores, many of these RTL designs still contain synthesizable logic beyond the instantiation of these unavailable modules. To use such designs in our work, we substitute unavailable IP cores with blank modules with appropriate interfaces to bypass the compilation errors. After that, we compile the Verilog code using Xcelium, employing HAL to generate a customized constraint file for the compilation process.   

The EDA compiler validates the synthesizability of RTL logic in each Verilog file, filtering out unsynthesizable designs. Meanwhile, the compiler also provides DFT-related information within the code including (1) the type of each DFT violation in the code, and (2) the specific line in which the DFT violation happens. Only four major types of DFT errors (ACNCPI, CLKNPI, CDFDAT, and FFCKNP, as discussed in the related works) are considered in this paper, while RTL designs containing other DFT error types are excluded from the dataset. For better focus, we also filtered out RTL designs containing multiple types of DFT violations, keeping those with single DFT error only.\footnote{As an initial effort in this field, we start from RTL designs with single type of DFT violation within the four critical types. Additional DFT error types, as well as more complex RTL designs with multiple DFT violations will be investigated in future works.}

\begin{figure}[htbp]
  \centering
  \subfloat[] {\label{fig:Dataset1} \includegraphics[width=0.20\textwidth]{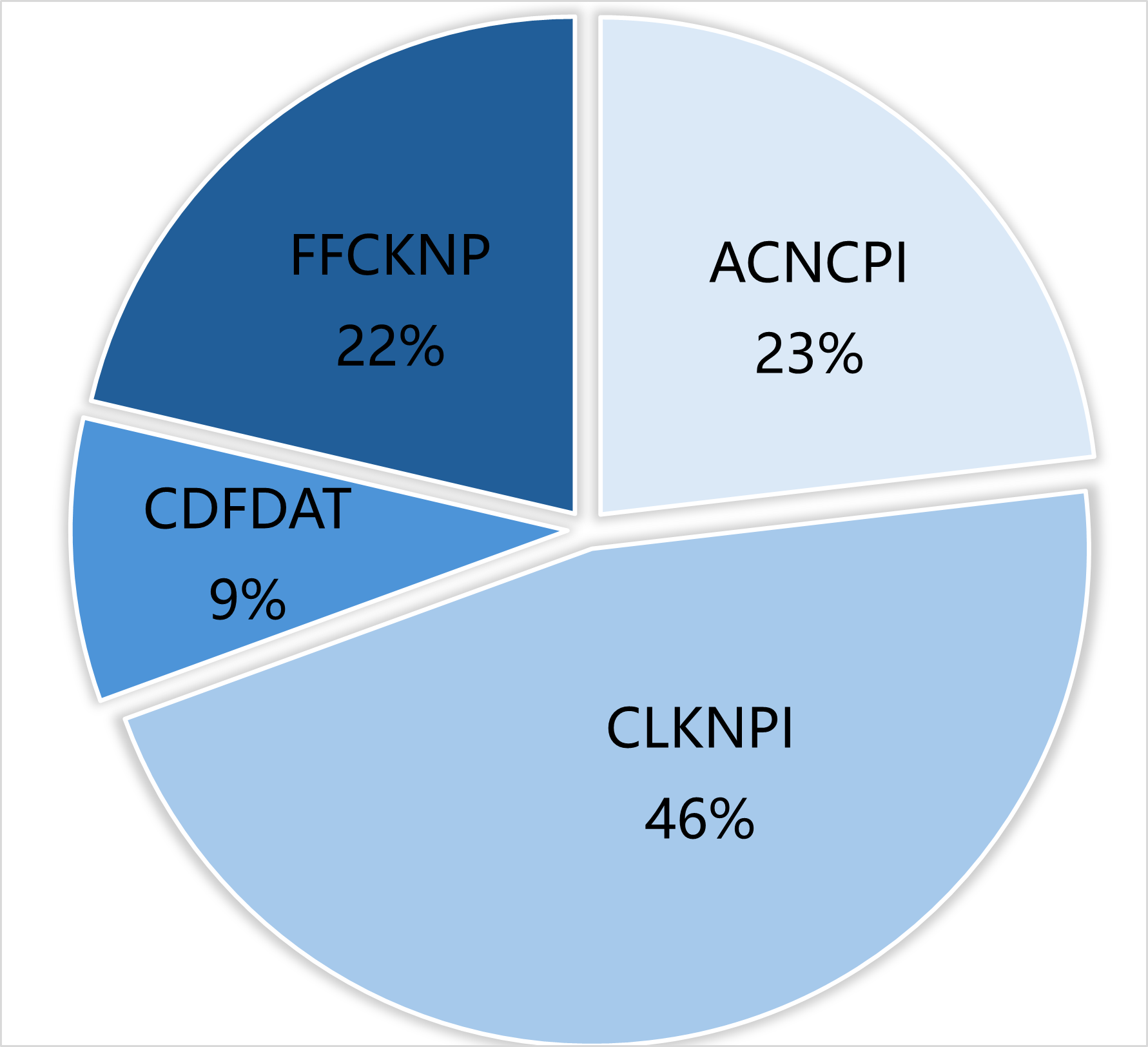}}
  \subfloat[] {\label{fig:Dataset2} \includegraphics[width=0.30\textwidth]{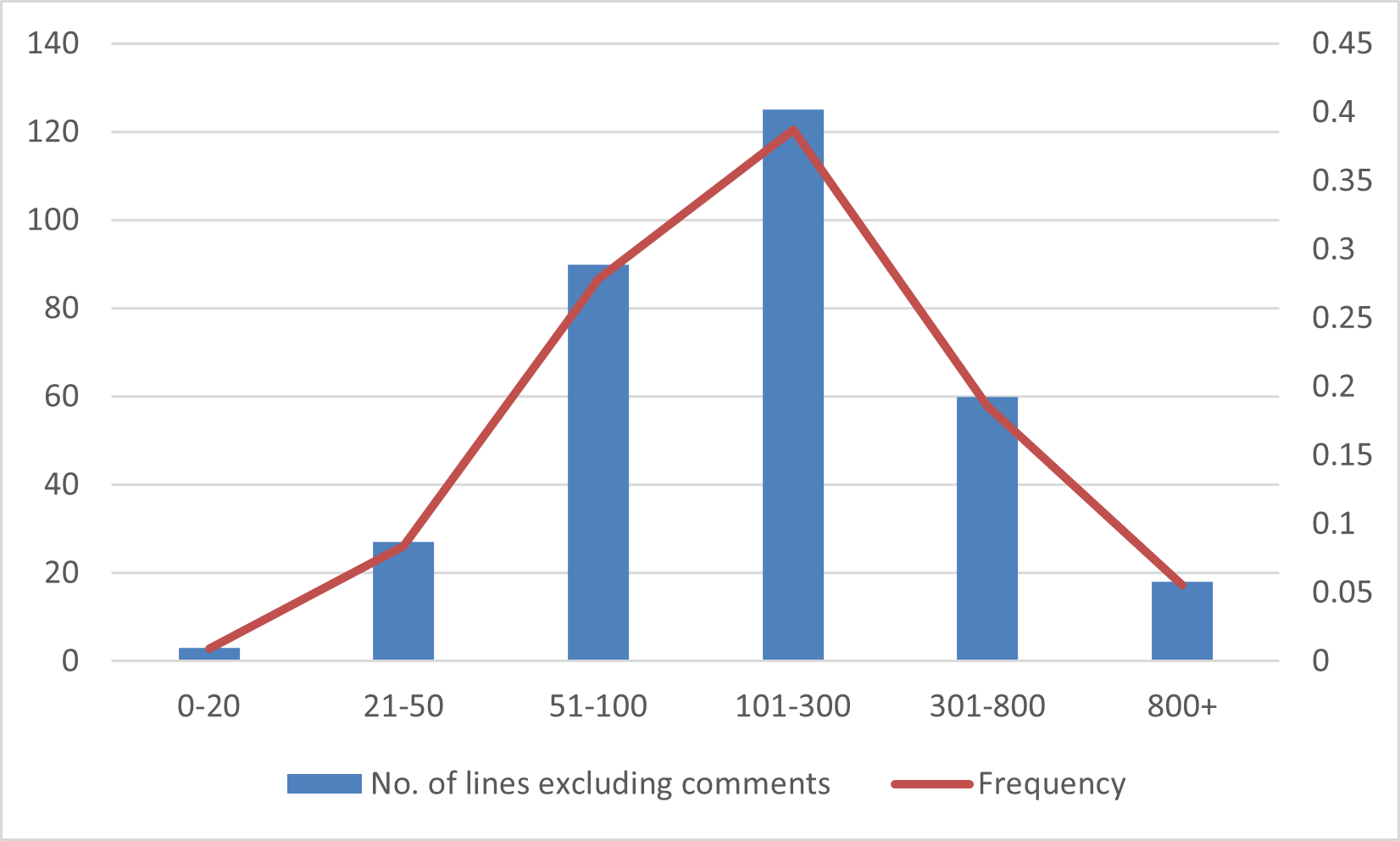}}
  \captionsetup{font={small}}
  \caption{Statistic overview of the VeriDFT dataset: (a) proportion of DFT-related error types, and (b) code length distribution histogram.}
\end{figure}

We obtained a total of 437 Verilog files with the above process. The proportion of DFT error types and the distribution of code length (in ``number of lines”) are given in Fig. \ref{fig:Dataset1} and Fig. \ref{fig:Dataset2}, respectively. 

The 437 files in VeriDFT are partitioned as follows. A relatively small training set is used to capture key characterization patterns. Most of the designs are reserved for rigorous testing of the framework.

\textbf{Training} (see Section \ref{sec:methodA}): Approximately 20\% (85 files) are used to train the autoencoder network using the contrastive learning approach. These training files are selected to encompass all DFT error types, ensuring the model encountered representative examples of each category. 

\textbf{Testing} (see Section \ref{sec:methodB}): (i) Reference Set  -- Approximately 8\% (35 files) are designated as the reference dataset during testing. Each RTL design in this reference set is paired with a validated revision – the revised version has been compiled with Xcelium and verified by Cadence Conformal LEC to ensure logical equivalence and DFT compliance. This curated set serves as a fixed library of retrieval examples during the RAG code repair process. (ii) Test Set  -- The remaining 72\% of the dataset (317 files) are used as the testing dataset to evaluate the performance of VeriRAG. During testing, an input RTL design from the test set and each reference sample from the reference set are passed through the autoencoder to generate their embedding vectors. The cosine similarity between the two embedding vectors is computed to identify the similarity between the input RTL design and the reference design.    After the similarity measurement, the most relevant reference RTL sample, paired with the associated manual revision, is retrieved as the LLM’s RAG reference to revise the given input RTL design to ensure DFT compliance.

\subsection{Verilog-JSON Conversion to Capture Hardware Similarity} \label{sec:DatasetB}
This subsection explains our motivation in using an EDA-generated JSON file for representing the hardware structure in an RTL design and introduces how the Verilog–JSON conversion is conducted. 

\begin{figure}[htbp] 
  \centering
  {\includegraphics[width=0.4\textwidth]{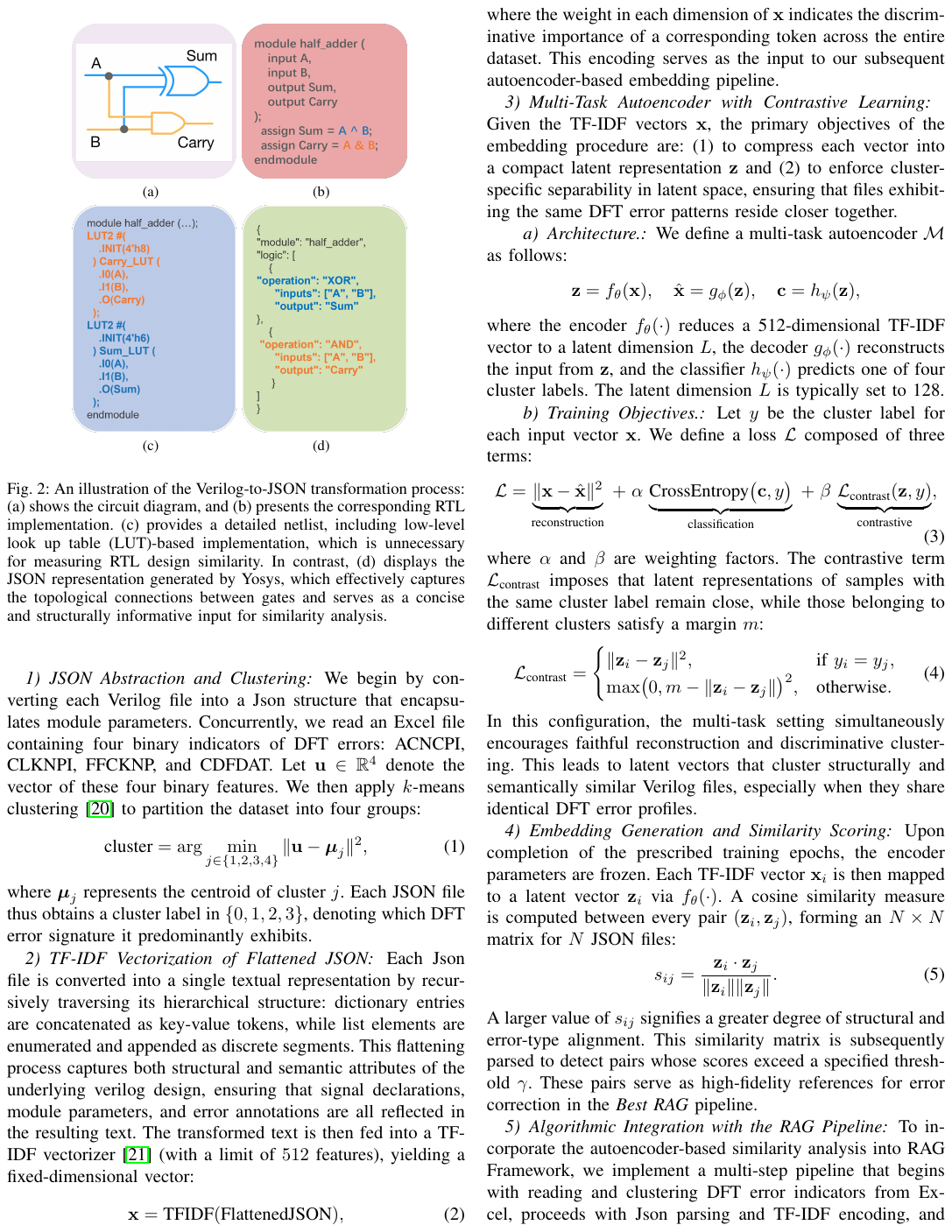}}
  \captionsetup{font={small}}
  \caption{Verilog-to-JSON transformation: (a) circuit diagram; (b) corresponding RTL implementation; (c) detailed netlist with low-level look up table (LUT) implementation (a distraction for similarity measurement); (d) JSON representation generated by Yosys capturing topological connections between major logic gates and sequential elements. In this work, “hardware structure” refers to these gate-level topologies, reflecting the structural context critical to DFT error patterns.}
  \label{fig:dataset4567}
\end{figure}

As the input to the autoencoder network, an RTL design should first be converted into a representation that retains essential information about the circuit's hardware structure while filtering out irrelevant implementation details. While Verilog codes and design netlists are straightforward representations that allow for textual similarity computations as described in prior works \cite{Ref18, Ref19}, they introduce DFT-irrelevant details that can distract the autoencoder. Specifically, Verilog code contains hardware-irrelevant information such as variable and parameter naming. Design netlists, on the other hand, include excessive low-level hardware details, such as the LUT-based implementation of gates. Fig. \ref{fig:dataset4567}(b) and Fig. \ref{fig:dataset4567}(c) give specific examples to illustrate the drawbacks of using Verilog code and design netlists for measuring design similarity.

To obtain the desired RTL representation, we transform Verilog designs into structured JSON files using Yosys, an open-source EDA synthesis tool. As illustrated in Fig. \ref{fig:dataset4567}(d), these JSON files are derived from netlists and emphasize topological connections among fundamental hardware structures (e.g., combinational/sequential logic, clock/data paths) instead of their implementation details (e.g., how the gate is realized with a look up table). This transformation captures the circuit topology while eliminating extraneous details.

As an important data pre-processing step before the model training and testing, we convert all RTL designs in VeriDFT into their JSON representations.

\section{Methodology}
\label{sec:Methodology_Section}

\begin{figure*}[htbp]
    \centering
    \includegraphics[width=1\linewidth]{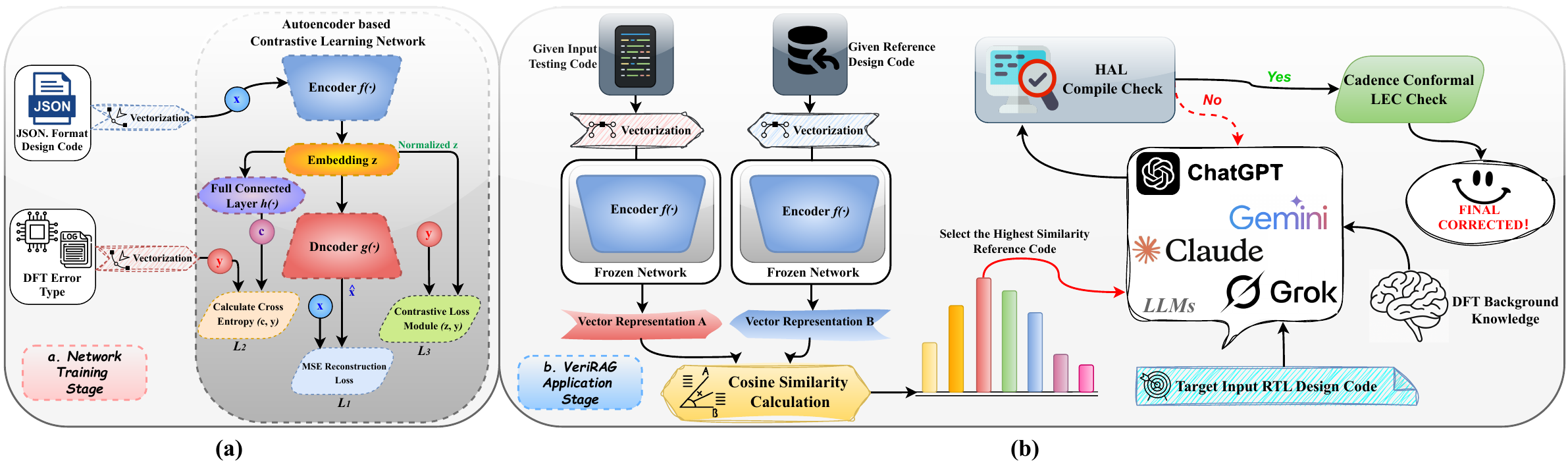}
    \captionsetup{font={small}}
    \caption{The VeriRAG framework -- (a) training of autoencoder network, (b) RAG-based code revision pipeline in the testing process.}
    \label{fig:verirag_flow}
\end{figure*}

This section presents the implementation of VeriRAG. Fig. \ref{fig:verirag_flow}(a) pertains to the training of the autoencoder, and Fig. \ref{fig:verirag_flow}(b) pertains to the overall application, where the autoencoder is used at the front end to identify the reference RTL design most similar to a given input RTL design. The most-similar reference design provides the reference for the subsequent LLM to perform code repair on the given input RTL design to ensure DFT compliance. 

\subsection{Training Autoencoder Network} \label{sec:methodA}
VeriRAG evaluates the similarity of RTL designs on two key aspects: (1) type of DFT violation, and (2) hardware structure. The preceding Section \ref{sec:DatasetA} explained how we obtain the DFT-error type labels that denote the type of DFT violation in an RTL design by analyzing the compilation log. Meanwhile, Section \ref{sec:DatasetB} detailed how we generate the JSON representation of an RTL design with Yosys. 

This section presents the training process for the autoencoder network, which maps an RTL design into a vector representation that intrinsically captures both (1) and (2) aspects of the design. As mentioned in Section \ref{sec:DatasetA}, the similarity between two RTL designs is quantified using the cosine similarity of their corresponding vector representations from the autoencoder outputs.

To this end, we train a multi-task encoder-decoder network with the training data in VeriDFT (Fig.\ref{fig:verirag_flow} (a)). First, we vectorize JSON representations and DFT error labels of the RTL designs in the training dataset. For the former, we apply the Term Frequency - Inverse Document Frequency (TF-IDF) vectorization scheme to convert a JSON description into numerical vector $\mathbf{x}$. This vectorization method has been proven to be effective for capturing structural patterns in hardware representations~\cite{thakur2023benchmarking, chen2023chemist}. For the implementation of TF-IDF encoder, we use the default smoothing and sublinear term frequency scaling provided by the Python scikit-learn toolbox, and we set $\mathbf{x}$ to be a 512-dimensional vector. For the latter, we construct a four-dimensional binary vector $\mathbf{y}$ with a simple label encoding method, using each binary of the vector to denote the existence of one type of DFT error, i.e., $y(i) = 0$ means the absence of DFT error type \#\textit{i} in the RTL design, while $y(i) = 1$ means the existence of DFT error type \#\textit{i} in the RTL design. Note for the dataset construction in Section~III that this paper, as an initial effort in the field, considers RTL designs with a single type of DFT violation, i.e., $\mathbf{y}$ is a one-hot vector in essence.

The training process follows the multi-task encoder-decoder training pipeline as in~\cite{Ref23}, in which an encoder $f(\cdot)$, a decoder $g(\cdot)$, and a DFT-error classifier $h(\cdot)$ are optimized together under a single joint loss (see Fig.\ref{fig:verirag_flow} (a) for the interconnectivities between the encoder, decoder and the classifier).

The overall loss is defined per batch of size $B$ as the weighted sum of a reconstruction term $L_1$, a classification term $L_2$, and a contrastive term $L_3$. Note that the decoder and classifier serve only as auxiliary supervisors to ensure the encoder performs good feature mappings in its vectorization process. Therefore, this paper does not present the training details of decoder and classifier. We refer interested readers to Section~III of~\cite{ref24} for a detailed introduction of the individual loss components used for decoder and classifier. Before presenting the overall loss function, we first elaborate on the three loss terms within the loss function below:

For the $k^\text{th}$ sample of batched inputs, the encoder takes the 512-dimensional numerical vector $\mathbf{x}_k$ as input and generates a 128-dimensional embedding vector output $\mathbf{z}_k$, i.e., $\mathbf{z}_k = f(\mathbf{x}_k)$. The decoder takes $\mathbf{z}_k$ as input and generates the reconstructed vector $\hat{\mathbf{x}}_k$, i.e., $\hat{\mathbf{x}}_k = g(\mathbf{z}_k)$. The Mean Squared Error (MSE) reconstruction loss $L_1$ ensures that the embedding vector faithfully preserves all details within the input JSON representation:
\begin{equation}
L_1 = \frac{1}{B} \sum_{k=1}^{B} \left\| \mathbf{x}_k - \hat{\mathbf{x}}_k \right\|_2^2
\end{equation}

We also want the encoder output $\mathbf{z}_k$ to capture the feature about the DFT error violation within the associated RTL design. To this end, we introduce the classification loss term in the general loss function. Specifically, $\mathbf{z}_k$ is fed to a classification network implemented with a fully connected layer. The output of the classification network is a 4-dimensional vector $\mathbf{c}_k = h(\mathbf{z}_k)$. The classification loss term $\mathcal{L}_2$ is defined as the cross entropy between $\mathbf{c}_k$ and the associated DFT error label $\mathbf{y}_k$, computed per sample and averaged across the same batch of size:
\begin{equation}
\mathcal{L}_2 = -\frac{1}{B} \sum_{k=1}^{B} \sum_{i=1}^{4} y_k(i) \log P_k(i)
\end{equation}

where $\mathbf{P}_k$ is the post-softmax probability vector, and $P_k(i)$ can be written as:
\begin{equation}
P_k(i) = \frac{\exp(c_k(i))}{\sum_{j=1}^{4} \exp(c_k(j))}
\end{equation}

We further introduce a contrastive term in the loss function to encourage embeddings of similar DFT error labels to be closer while pushing those with significantly different labels apart so that RTL designs with the same DFT error types are likely to have embedding high similarity measure, and vice versa. Following the classic setup in~\cite{Ref23}, given a batch of normalized embeddings $\{\mathbf{z}_k^{\text{norm}}\}_{k=1}^{B}$, in which we have $\mathbf{z}_k^{\text{norm}} = \mathbf{z}_k / \|\mathbf{z}_k\|$, and the associated batch of label vector $\{\mathbf{y}_k\}_{k=1}^{B}$, we have:
\begin{equation}
\mathcal{L}_3 = \frac{1}{N} \sum_{p < q}
\begin{cases}
\left\| \mathbf{z}_p^{\text{norm}} - \mathbf{z}_q^{\text{norm}} \right\|^2, & \mathbf{y}_p = \mathbf{y}_q \\
\left[ \max \left( m - \left\| \mathbf{z}_p^{\text{norm}} - \mathbf{z}_q^{\text{norm}} \right\|, 0 \right) \right]^2, & \mathbf{y}_p \ne \mathbf{y}_q
\end{cases}
\end{equation}

where $m$ is a hyperparameter representing the fixed margin, describing the targeted minimum distance between two normalized embedding pairs with different labels. $N$ denotes the total number of unique unordered sample-pairs in the batch, i.e. $N = B(B - 1)/2$.

Finally, with the above background, we give the expression of the loss function as
\begin{equation}
\mathcal{L} = \mathcal{L}_1 + \alpha \mathcal{L}_2 + \beta \mathcal{L}_3
\end{equation}
where we have $\alpha = 0.01$ and $\beta = 0.01$ in our implementation.

\begin{figure*}[!t]
  \centering
  % Replace filename if needed, e.g., rate_pretty_preliminary_row5_v3.pdf
  \includegraphics[width=\textwidth]{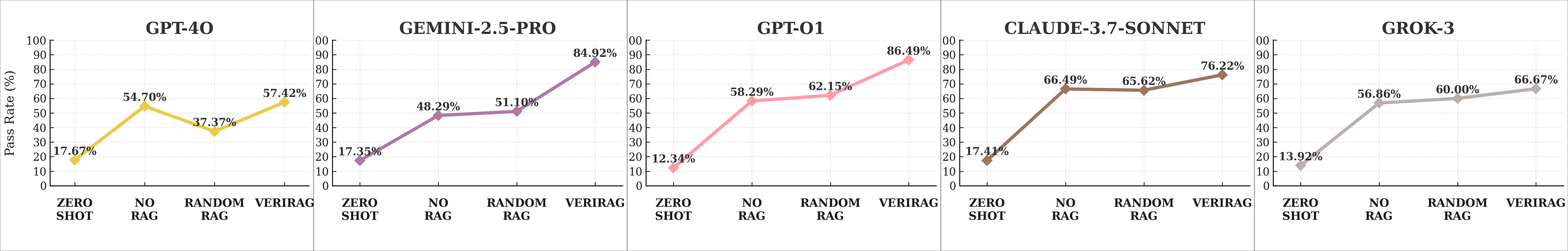}
  \captionsetup{skip=4pt} % tighten space between figure and caption (local)
  \caption{Success rates of preliminary DFT error corrections (without considering logical equivalence) for different LLMs and ablation cases tested.}
  \label{fig:success_rates_1}
\end{figure*}

\subsection{Application of VeriRAG framework } \label{sec:methodB}
The above training process ensures that the encoder network can well identify the hardware structure and DFT violation of a given RTL design input and properly encodes the above information into the embedding vector. This subsection concerns the application of the encoder in the overall VeriRAG framework.

Given the JSON representation of an input RTL design, denote the associated embedding vector generated by the encoder network by $\mathbf{z}^T$, where the superscript $T$ indicates that this is a testing input. Further, denote the embedding vector for reference RTL design $r$ in the reference dataset by $\mathbf{z}_r^R$, where the superscript $R$ indicates that this is a reference sample. The similarity between the given RTL design and reference RTL design $r$, denoted as $s_r$, is defined as the cosine similarity between $\mathbf{z}^T$ and $\mathbf{z}_r^R$, i.e.,
\begin{equation}
s_r = \frac{\mathbf{z}^{T} \cdot \mathbf{z}_r^R}{\left\| \mathbf{z}^{T} \right\| \left\| \mathbf{z}_r^R \right\|}, \quad \text{where } \mathbf{z}^{T} = f(\mathbf{x}^{T}),\; \mathbf{z}_r^R = f(\mathbf{x}_r^R)
\end{equation}

After comparing the input RTL design and all reference RTL designs in the reference dataset, we obtain a vector of similarity score, i.e., $\mathbf{s} = \langle s_1, s_2, \dots, s_n \rangle$, where $n$ is the number of reference RTL designs within the reference dataset. By traversing this similarity vector $\mathbf{s}$, we select the reference RTL design with the highest similarity score ($s_{\max}$) as the most relevant reference. We refer to the combination of this reference design and its manually edited revision as the ``reference-answer pair''.

The VeriRAG framework leverages this reference-answer pair retrieved from the reference dataset to iteratively refine the target input Verilog code. Along with the reference-answer pair, the initial prompt given to the LLM -- which is responsible for generating the DFT-compliance code -- also contains (1) a general description of the task, (2) a background introduction on testability principles in RTL design, and (3) detailed definitions of the four DFT error types being considered.

Using these prompt inputs, the LLM generates an initial fix for the target Verilog code. To ensure the revised Verilog code is compilable and DFT-compliant, we introduce a feedback loop to check the synthesizability and DFT violation of the resulting design. If any compilation error or DFT violation is found by the EDA compiler (Xcelium, in our implementation), the compilation log will be provided to the LLM, which uses this feedback to refine the code in the next iteration. This process continues until a successful compilation is achieved or the maximum number of iterations ($K$, set to 5 in our implementation) is reached.

Once a successful revision passes the compilation check, it undergoes a final Cadence Conformal Logic Equivalence Check (LEC) to validate logical equivalence. Upon passing this rigorous validation, the RTL design is marked as correctly revised, concluding a successful iterative revision process.

\section{Experiments}
\label{sec:Experiments}

\begin{figure*}[!t]
  \centering
  % Replace filename if needed, e.g., rate_pretty_ultimate_row5_v3.pdf
  \includegraphics[width=\textwidth]{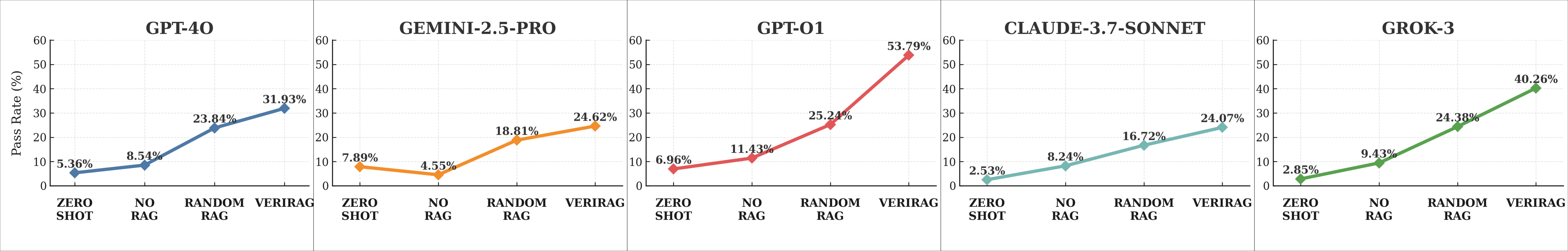}
  \captionsetup{skip=4pt} % tighten space between figure and caption (local)
  \caption{Ultimate code revision success rates (with logical equivalence) for different LLMs and ablation cases tested.}
  \label{fig:success_rates_2}
\end{figure*}
This section evaluates VeriRAG by comparing it against three ablation variants described below. The goal of these experiments is to assess the contribution of each component within VeriRAG to the DFT error correction.

\textbf{Zero Shot:} The tested LLM receives only the target Verilog file containing DFT errors without any hints -- no background information or RTL design references are included in the prompt. This straightforward baseline measures the generic LLM’s ability to handle DFT errors in a single inference step without iterative feedback and error correction.

\textbf{No RAG:} In addition to the RTL design requiring revision, the LLM is provided with general DFT knowledge and brief descriptions of the four specific DFT error types. This background information allows the LLM to better understand the task. However, no reference RTL design or iterative refinement process is involved.

\textbf{Random RAG:} Along with the general DFT background, the LLM is given a Verilog sample randomly selected from the reference dataset, paired with its corresponding revision. The random reference design may not match the structure or error type of the target RTL design. We adopted the aforementioned iterative revision policy for Random RAG: for fairness, both VeriRAG and Random RAG are allowed up to five error-correction attempts. In each iteration, the LLM receives the most recent corrected version of the RTL (from the previous round), along with its compiler diagnostics. Although the reference sample remains unchanged, the input context evolves iteratively based on the updated design and feedback, enabling progressive refinement over iterations. This is the case for both Random RAG and VeriRAG. The only difference between this baseline and our method is in the retrieval policy -- VeriRAG uses the autoencoder to select the most relevant example, while Random RAG selects references randomly.

As mentioned in Section~\ref{sec:Dataset}, the VeriDFT dataset comprises 35 reference RTL designs (each paired with a standard revision) and 317 test files. In the following experiments, we tested all RTL designs in the testing dataset to enable comprehensive assessment using the reference dataset as the retrieval knowledge base. Our experiments focus on two main questions: (1) What proportion of the generated RTL revisions can fully eliminate DFT error while remaining synthesizable? (2) Among those revisions satisfying 1), what proportion also preserves logical equivalence with the original design?

Here, (2) is a subset of (1). Both DFT error elimination and synthesizability in (1) are verified by the compiler, so we consider them together. The rest of this section addresses these questions by testing five state-of-the-art LLMs: GPT-o1, GPT-4o, Grok-3, Claude-3.7-Sonnet, and Gemini-2.5-Pro.

We next present experimental results for the tested LLMs under different prompting approaches. Given the test dataset, let $ANS$ be the total number of RTL revisions generated. Let $NoErr$ denote the number of RTL revisions that are both free of DFT errors and remain synthesizable, corresponding to question (1) but not (2). Fig.~\ref{fig:success_rates_1} shows the preliminary DFT error correction rate ($NoErr / ANS$) across different LLMs and ablation variants. For these error-free RTL revisions, we further conducted LEC. Let $Eq$ be the number of logically equivalent RTL designs, corresponding to satisfying both question (1) and (2). Fig.~\ref{fig:success_rates_2} presents the ultimate success rate ($Eq / ANS$) for different LLMs and all ablation cases.

Three key trends are evident across these experimental results. First, the incremental integration of domain-specific context and structured references consistently improves the LLM’s abilities to correct DFT errors. Transitioning from a Zero Shot setting to the advanced VeriRAG configuration increases the ultimate success rates for all tested LLMs. These findings support our central thesis: precise, structure-aware RTL references, combined with the iterative code revision pipeline, significantly outperform generic textual prompts or random reference examples.

Second, the results reveal notable variation in how different LLMs utilize incremental guidance. Grok-3 achieves the largest relative gain, indicating its superior utilization of structural references and iterative feedback, with its logic equivalence rate rising from a minimal 2.85\% with Zero Shot to 40.26\% with VeriRAG. GPT-4o also benefits markedly, reaching 31.93\% equivalence, while Claude-3.7-Sonnet and Gemini-2.5-Pro plateau in the mid-20\% range. These results highlight that general purpose LLMs differ significantly in their intrinsic abilities to integrate structured references and iterative compiler feedback into effective RTL corrections.

Finally, GPT-o1 provides a meaningful upper-bound reference, clearly illustrating the potential of large-scale, general-purpose LLMs in the VeriRAG framework. It achieves an ultimate success rate of 53.76\%, a 7.72-fold improvement over its Zero Shot baseline. This impressive margin emphasizes the variable win achieved with enhanced structural retrieval and precise compiler-guided iterations, even without domain-specific tuning.

\section{Conclusions and Outlook}
\label{sec:Conclusions}

We present \textit{VeriRAG}, a retrieval-augmented framework for repairing DFT errors at the RTL stage. Concretely, the approach employs an autoencoder trained with contrastive learning to retrieve structurally similar references from a newly developed Verilog DFT dataset, \textit{VeriDFT}. Leveraging these targeted references, an LLM performs repairs through an iterative Verilog code-revision loop guided by compiler and DFT diagnostics. A candidate fix is accepted only when the design remains synthesizable, contains no DFT violations, and passes a final logic equivalence check (LEC), ensuring that improvements in testability do not compromise functional intent.

On a benchmark of 317 RTL designs, \textit{VeriRAG} improves the success rate across several state-of-the-art LLMs—for example, from 3\% to 40.26\% with Grok-3 and from 6.96\% to 53.76\% with GPT-o1 (13.42$\times$ and 7.72$\times$ improvements, respectively). Ablation studies further indicate that iteration alone is insufficient and that structure-aware retrieval is essential; results improve consistently from Zero-shot to No-RAG to Random-RAG to \textit{VeriRAG}. Taken together, these findings show that structurally aligned retrieval and tool-guided iteration act as complementary levers for reliable RTL-level DFT repair, yielding substantive gains without modifying downstream tool flows.

This study has practical boundaries. The dataset focuses on single-violation cases across four representative DFT types, reflecting a controlled setting for isolating effects. Blank-module substitution is applied only when interfaces and connectivity are preserved; cases that would suppress or fabricate diagnostics are excluded, though residual bias cannot be completely ruled out. The reference library is intentionally modest in size to stress retrieval quality and to make failure modes more interpretable.

Looking forward, several directions appear promising. First, domain-adaptive pretraining or fine-tuning of LLMs on Verilog and DFT corpora could strengthen sensitivity to logical-equivalence constraints \cite{qi2025governanceawarehybridfinetuningmultilingual,Qi_2025}. Second, lightweight formal-in-the-loop checks (SEC/LEC) at each iteration may convert equivalence gaps into actionable feedback, narrowing the distance between preliminary and ultimate success \cite{cui2024origen}. Third, moving beyond static Verilog-to-JSON vectorization toward learnable (de)vectorizers that preserve logic-level invariants could improve both retrieval fidelity and edit quality\cite{qi2026topoedgetopologygroundedagenticframework}. Finally, expanding the dataset to cover additional DFT types, multi-violation designs, and a larger, more diverse set of validated references would better stress-test retrieval at scale and sharpen external validity. Collectively, these extensions position \textit{VeriRAG} as a practical foundation for integrating LLMs into DFT-aware RTL workflows.

%\section{Acknowledgment}
%This work was supported in part by the Shenzhen-Hong Kong-Macao technical program under Grant No. SGDX20230821094359004.

%\input{main.bbl}
{\footnotesize
\bibliographystyle{IEEEtran}
\bibliography{reference}
}

\end{document}